\documentclass[aps,preprint]{revtex4}%
\usepackage{amsfonts}
\usepackage{amsmath}
\usepackage{amssymb}
\usepackage{graphicx}%
\begin{document}
\preprint{ }
\title[Short title for running header]{Chaos in Kundt type III Spacetimes}
\author{I.Sakalli}
\email{izzet.sakalli@emu.edu.tr}
\author{M.Halilsoy}
\email{mustafa.halilsoy@emu.edu.tr}
\affiliation{Physics Department, Eastern Mediterranean University, G.Magosa, N.Cyprus,
Mersin 10, Turkey.}

\begin{abstract}
We consider geodesics motion in a particular Kundt type III spacetime in which
Einstein-Yang-Mills equations admit solutions. On a particular surface as
constraint we project the geodesics into the $(x,y)$ plane and treat the
problem as a 2-dimensional one. Our numerical study shows that chaotic
behavior emerges under reasonable conditions.

\end{abstract}
\volumeyear{year}
\volumenumber{number}
\issuenumber{number}
\eid{identifier}
\date[Date text]{date}
\received[Received text]{date}

\revised[Revised text]{date}

\accepted[Accepted text]{date}

\published[Published text]{date}

\maketitle

Kundt's class of solutions present spacetimes with non-expanding, shear-free
and twist-free null geodesic congruences \cite{R1}. Interest in this class of
solutions comes from the fact that these spacetimes admit plane waves, which
exhibit geometrically different properties than the \textit{pp}-waves. Due to
its potential application in string theory Kundt class still maintains its
popularity \cite{R2,R3,R4,R5,R6}. Various Petrov types of Kundt solutions were
identified, among them especially Kundt type N has been studied in detail for
long time by many authors,(\textit{see for instance} \cite{R7,R8,R9}).
Geodesics in a specific Kundt type N was analyzed in detail by \cite{R9} , and
it was shown that particular solutions obey a power-law. Recently, interest in
the Kundt type III spaces has gained momentum. Firstly, Griffiths \textit{et
al} \cite{R10} derived and classified a complete family of Kundt type III,
which admit cosmological constant $\Lambda_{c}$\ and/or pure radiation
$\Phi22\neq0$. In the absence of the cosmological constant $\Lambda_{c}$, the
Kundt type III solutions are further generalized to be the solutions of the
Einstein-Yang-Mills (EYM) system by Fuster and Holten \cite{R11}.

The pioneering study of proving chaos in the spacetimes of plane waves was
done for pure impulsive gravitational \textit{pp}-waves \cite{R12} . Recently,
it has been further demonstrated that under certain conditions the emergence
of chaotic motion is possible both in the spacetimes of the superposed
electrovac \textit{pp}-waves and in the non-Abelian plane waves of Kundt type
N \cite{R13}, which are the solutions of \ the $D=4$ EYM equations. In
\cite{R13}, it was also pointed out that for the non-Abelian plane waves the
chaotic effect of gravity dominates over the gauge field. This is due to the
fact that such local fields vanish asymptotically and the chaos inherited from
gravity renders the whole system chaotic.

Analysis of the geodesic motion for the $D=4$ Kundt type III is also discussed
by \cite{R11} in which a possible chaotic motion is highlighted in particular
cases of this class.

In this Brief Report, our aim is to investigate the possibility of chaotic
geodesics in Kundt class III spacetimes. For this purpose, we consider the
following algebraically special line-element \cite{R3,R11},%

\begin{equation}
ds^{2}=2du[dv+Hdu+Wdz+\overline{W}d\overline{z}]-2dzd\overline{z},
\end{equation}

\bigskip where $H=H(u,v,z,\overline{z})$ is a real function while
$W=W(u,v,z,\overline{z})$ is a complex function, in general. Here our
motivation is to study the geodesic motion for the case $W_{,v}=\frac
{-2}{z+\overline{z}}$, referring to the solution of the EYM equations for the
metric (1) given by\ \cite{R11}%

\[
W=W^{0}(u,z)-\frac{2v}{z+\overline{z}},
\]

\begin{equation}
H=H^{0}(u,z,\overline{z})+\frac{W^{0}(u,z)+\overline{W}^{0}(u,\overline{z}%
)}{z+\overline{z}}v-\frac{v^{2}}{(z+\overline{z})^{2}}.
\end{equation}

where $W^{0}(u,z)$ is an arbitrary complex function and $H^{0}(u,z,\overline
{z})$ is a real function. The simplest choice of $W^{0}(u,z)$ for the type III is%

\begin{equation}
W^{0}(u,z)=g(u)z,
\end{equation}

such that the Weyl scalar $\Psi_{3}\neq0.$ On the other hand, imposing the
only solution ($\chi^{a}=\lambda^{a}(u)z$) on the Yang-Mills (YM) equation in
which the energy density is bounded throughout the spacetime, the solution for
$H^{0}(u,z,\overline{z})$ becomes%

\begin{equation}
H^{0}(u,z,\overline{z})=\left[  f(u,z)+\overline{f}(u,\overline{z})\right]
(z+\overline{z})-g\overline{g}z\overline{z}+\sigma(u)(z+\overline{z}%
)^{2}\left[  \ln(z+\overline{z})-1\right]  .
\end{equation}

where $f(u,z)$ is an arbitrary complex function and $\sigma(u)=2\gamma
_{ab}\lambda^{a}(u)\overline{\lambda}^{b}(u)+g(u)\overline{g}(u)$ is a real
function. Here, $\lambda^{a}(u)$ are bounded complex functions and
$\gamma_{ab}$ is the invariant metric of the Lie group. We note that the
condition for being Kundt type III spacetimes is $g(u)\neq0$. It is trivially
seen that $\lambda^{a}(u)=0$ corresponds to the vacuum solution.

Our primary interest here is to write the geodesics equations for the metric
(1). Similar to the study \cite{R11}, for the beginning, we eliminate the
$u$-dependence from \ $W^{0}$ and $g$ by the following particular choice%
\begin{equation}
W^{0}=z\text{ \ \ \ \ \ and \ \ \ \ }g=1,
\end{equation}

Next, introducing real spatial coordinates $x$ and $y$ by $\sqrt{2}z=\left(
x+iy\right)  $, we get the geodesic equations as%

\begin{equation}
\overset{\cdot\cdot}{u}-\overset{\cdot}{u}^{2}(1-\frac{v}{x^{2}}%
)+2\overset{\cdot}{u}\frac{\overset{\cdot}{x}}{x}=0,
\end{equation}

\begin{equation}
\overset{\cdot\cdot}{x}+\overset{\cdot}{u}^{2}\left[  H_{,x}-\left(
1-\frac{v}{x^{2}}\right)  \left(  x-2\frac{v}{x}\right)  \right]  +\frac{2}%
{x}\overset{\cdot}{u}\overset{\cdot}{v}+2\overset{\cdot}{u}\overset{\cdot}%
{x}\left(  1-2\frac{v}{x^{2}}\right)  =0,
\end{equation}

\begin{equation}
\overset{\cdot\cdot}{y}+\overset{\cdot}{u}^{2}\left[  H_{,y}+y\left(
1-\frac{v}{x^{2}}\right)  \right]  -2\overset{\cdot}{u}\overset{\cdot}{x}%
\frac{y}{x}=0,
\end{equation}

where the dot denotes $\frac{d}{d\tau}$ with $\tau$ being the proper time. In
addition the metric condition implies%

\begin{equation}
\overset{\cdot}{x}^{2}+\overset{\cdot}{y}^{2}-2\overset{\cdot}{u}%
\overset{\cdot}{v}-2H\overset{\cdot}{u}^{2}-2\overset{\cdot}{u}\overset{\cdot
}{x}(x-2\frac{v}{x})+2y\overset{\cdot}{u}\overset{\cdot}{y}=\epsilon.
\end{equation}

where $\epsilon=1,0,-1$ stands for timelike, null and spacelike geodesics,
respectively. The present form of the equation set does not allow us to obtain
a 2D $(x,y)$ Hamiltonian system analogous to the previous studies
\cite{R12,R13}. However, with appropriate choices of $u$ and $v$ surfaces it
is possible to project the geodesics into the $(x,y)$ plane in which writing a
2D Hamiltonian becomes possible. Our first intention is to shift the
independent variable from $\tau$ to $u$ as an affine parameter.\ If we
consider a family of geodesics, which follow the light-cone coordinate $u$
with constant rate of change in the same proper time intervals, the following
assumption can be made%

\begin{equation}
\overset{\cdot}{u}=\text{constant}\equiv1,
\end{equation}

Such an assumption gives rise to a condition on the $v$ surfaces given by%

\begin{equation}
v=x^{2}(1-2\frac{x^{\prime}}{x}),
\end{equation}

Here " $\overset{\prime}{}$ " denotes $\frac{d}{du}$. By this substitution
into Eqs. (7) and (8), we get a 2D dynamical system in the ($x,y$) plane%

\begin{equation}
3x^{\prime\prime}-H_{,x}^{0}-x=0,
\end{equation}

\begin{equation}
y^{\prime\prime}+H_{,y}^{0}=0,
\end{equation}

which is described by a Super-Hamiltonian \cite{R14}%

\begin{equation}
\mathcal{H=}\frac{1}{2}\left(  P_{y}^{2}-\frac{P_{x}^{2}}{3}\right)  +V(x,y),
\end{equation}

with the corresponding potential%

\begin{equation}
V(x,y)=H^{0}+\frac{x^{2}}{2}.
\end{equation}

Let us note that the Super-Hamiltonian defined by the momenta $P_{x}%
=-3x^{\prime}$\ and $P_{y}=y^{\prime}$ is not positive definite.

Eq. (9) stands for an energy condition in which it should be automatically
satisfied by the solutions of Eqs. (12) and (13). Without loss of generality,
we can assume that $f$ and $\lambda^{a}$ are independent of $u$. This
assumption implies that $\sigma$ is a positive constant. As we mentioned
before that the chaotic effect of gravity dominates over the gauge (YM) field
\cite{R13}, it would be sufficient to investigate chaos in vacuum, i.e.
$\sigma=1.$ In other words, once the chaotic motion appears in the vacuum
spacetime, the local fields could not be strong enough to avert it into a
regular motion.

In general, any $f=kz^{n}$ ($n=0,1,2...$)$,$ with the multiplicative factor
$k$ being an arbitrary real parameter implies a potential%

\begin{equation}
V=\sqrt{2}kx\operatorname{Re}(z^{n})-\frac{y^{2}}{2}+2x^{2}\left[  \ln
(\sqrt{2}x)-1\right]  .
\end{equation}

which admits an integrable system for $k=0.$ The logarithmic term in the
potential imposes a condition on the $x$ coordinate, namely $x>0$. Beside
this, the case $f=kz$, which describes a flat space for vacuum \textit{pp}%
-wave spacetimes \cite{R1,R12}, confesses a regular motion for the geodesics
particles. Contrary to the vacuum homogeneous \textit{pp}-waves \cite{R12} ,
here the case $n=2$ admits a nonintegrable dynamical system. However, the
nonintegrable systems in the vacuum \textit{pp}-waves were obtained for the
cases with $n\geq3,$ \cite{R12} .

Here we wish to study the nonintegrable system with the simplest case ($n=2$),
and explore whether the motion depends on the initial conditions or not. If
chaos emerges in such a simplest case, intuitively it should also appear for
the $f$ functions with higher powers of $z$. Now, for $n=2$, it can be seen
that the potential (16) has various unstable points according to the range of
the multiplicative constant $k$.

TABLE I: Unstable points depending on the value of $k$.%

\begin{tabular}
[c]{|c|c|c|}\hline
Points & Saddle & Repellor\\\hline
$\left\{  x=\frac{-1}{2\sqrt{2}k},\text{ }y=\pm\frac{1}{2\sqrt{2}k}%
\sqrt{7+8\ln(-2k)}\right\}  $ & $k<-\frac{1}{2}e^{-\frac{7}{8}}$ & ---\\\hline
$\left\{  x=\frac{2\sqrt{2}}{3k}LambertW\left(  \frac{3}{4}k\sqrt{e}\right)
,\text{ }y=0\right\}  $ & $k\geqslant-\frac{1}{2}e^{-\frac{7}{8}}$ &
$-\frac{4}{3}e^{-\frac{3}{2}}<k<-\frac{1}{2}e^{-\frac{7}{8}}$\\\hline
\end{tabular}

\bigskip Those results in the table show us that the highest possibility of
the emergence of chaos corresponds to the case\ $k<-\frac{1}{2}e^{-\frac{7}%
{8}}$ , in which admits two saddle points. Particularly, the case $-\frac
{4}{3}e^{-\frac{3}{2}}<k<-\frac{1}{2}e^{-\frac{7}{8}}$ has an additive
repellor point, and whence it may follow a stronger chaos. Conversely, the
case $k\geqslant-\frac{1}{2}e^{-\frac{7}{8}}$ upon possessing one saddle point
causes a questionable chaotic motion. In order to judge the existence of the
chaotic motion, we study the numerical analysis of the evolution of the test
particle in the gravitational field.

We integrate numerically the equations of motion given by Eqs. (12) and (13).
The initial conditions depend on 3 parameters, $\left(  x_{0},\text{ }%
y_{0}\right)  $ (at $u=0$) and $k$. For a given $k$ value, we may choose
$\left(  x_{0},\text{ }y_{0}\right)  $ such that keeping $x_{0}$ unchanged,
and checking the effect of the $y_{0}$ on the geodesic motions, while it
varies. To do this, we may set $x_{0}=c_{1}$, a real constant $c_{1}>0,$ and
$y_{0}=-3+\underset{j=0}{\overset{18}{\sum}}\frac{j}{3}$. For example, if we
take $k=0,$ the solutions are trivially analytic. This is also graphically
verified in Fig. 1. Next, by considering the cases $k\neq0$\ the motion can
lead to a chaotic motion. It is observed that the chaos has a \textit{movable
character} depending on the choices of $x_{0}$ and $y_{0}$ while
$k\geqslant-\frac{1}{2}e^{-\frac{7}{8}},$ see Fig. 2.\textit{ }However, when
the multiplicative constant $k<-\frac{1}{2}e^{-\frac{7}{8}}$ chaos is obvious.
It is seen that the multiplicative constant $k$\ of the function $f$ becomes
decisive for the chaotic motion. In other words, $k$ plays the role of
critical parameter for the onset of chaos. The chaotic behavior \ of the
geodesics in the case $k<-\frac{1}{2}e^{-\frac{7}{8}}$ is illustrated in Fig.
3. Alternatively, the dynamical system can be investigated by using the
Poincar\'{e} section method. We use the package POINCAR\'{E} \cite{R14} to
perform the numerical experiments. Fig. (4) is a demonstration of the
Poincar\'{e} section, which verifies the chaotic behavior in our dynamical system.

In conclusion, it is shown that the Kundt type III spacetimes may reveal
chaotic motion under certain conditions. To our knowledge such a study did not
exist in the literature before. Chaos in the spacetimes of electrovac and the
specific Kundt type N with YM field was studied before \cite{R13}. This report
constitutes an extension of that study. It is needless to state that the
existence of the chaos in the Kundt type III spacetimes may have further
implications for the particle motions in string theory and in higher dimensions.

\newpage

\begin{center}
Figures
\end{center}

FIG. 1: For $k=0$, 2D ($x,y$) plot of the geodesics. Geodesics start from
$\left\{  x_{0}=0.7,\text{ }y_{0}=-3+\underset{j=0}{\overset{18}{\sum}}%
\frac{j}{3}\right\}  $ (dashed line) and move through non-intersected
trajectories. The non-intersected trajectories represent the regular motion.

FIG.2: In case $k>-\frac{1}{2}e^{-\frac{7}{8}}$, $k=1$ is chosen for the 2D
($x,y$) plot of the geodesics.The initial positions are $\left\{
x_{0}=0.7,\text{ }y_{0}=-3+\underset{j=0}{\overset{18}{\sum}}\frac{j}%
{3}\right\}  $ (dashed line). Intersected trajectories signal the existence of
chaos. Two trajectories, which have different $y_{0}$ initial points initially
accelerate in $+x$-direction contrary to the others.

FIG. 3: In case $k<-\frac{1}{2}e^{-\frac{7}{8}}$, $k=-0.21$ $(-\frac{4}%
{3}e^{-\frac{3}{2}}<-0.21<-\frac{1}{2}e^{-\frac{7}{8}})$ is chosen 2D ($x,y$)
plot of the geodesics.The initial positions are $\left\{  x_{0}=0.7,\text{
}y_{0}=-3+\underset{j=0}{\overset{18}{\sum}}\frac{j}{3}\right\}  $ (dashed
line). The chaotic behavior is evident from the trajectories. The symmetry in
Eq. (13) shows itself along the $y$-axis.

FIG. 4: Poincar\'{e} sections of $(x^{\prime},x)$ for $k=-0.21$ and $H=0.2$
across $y=0$ KAM surface. Some points are distributed randomly in a finite
region to form a chaotic sea, however the large island surrounded by the
chaotic sea indicates the existence of quasi-periodic orbits. (Here,
$x\rightarrow q1,$ and $x^{\prime}\rightarrow p1$)

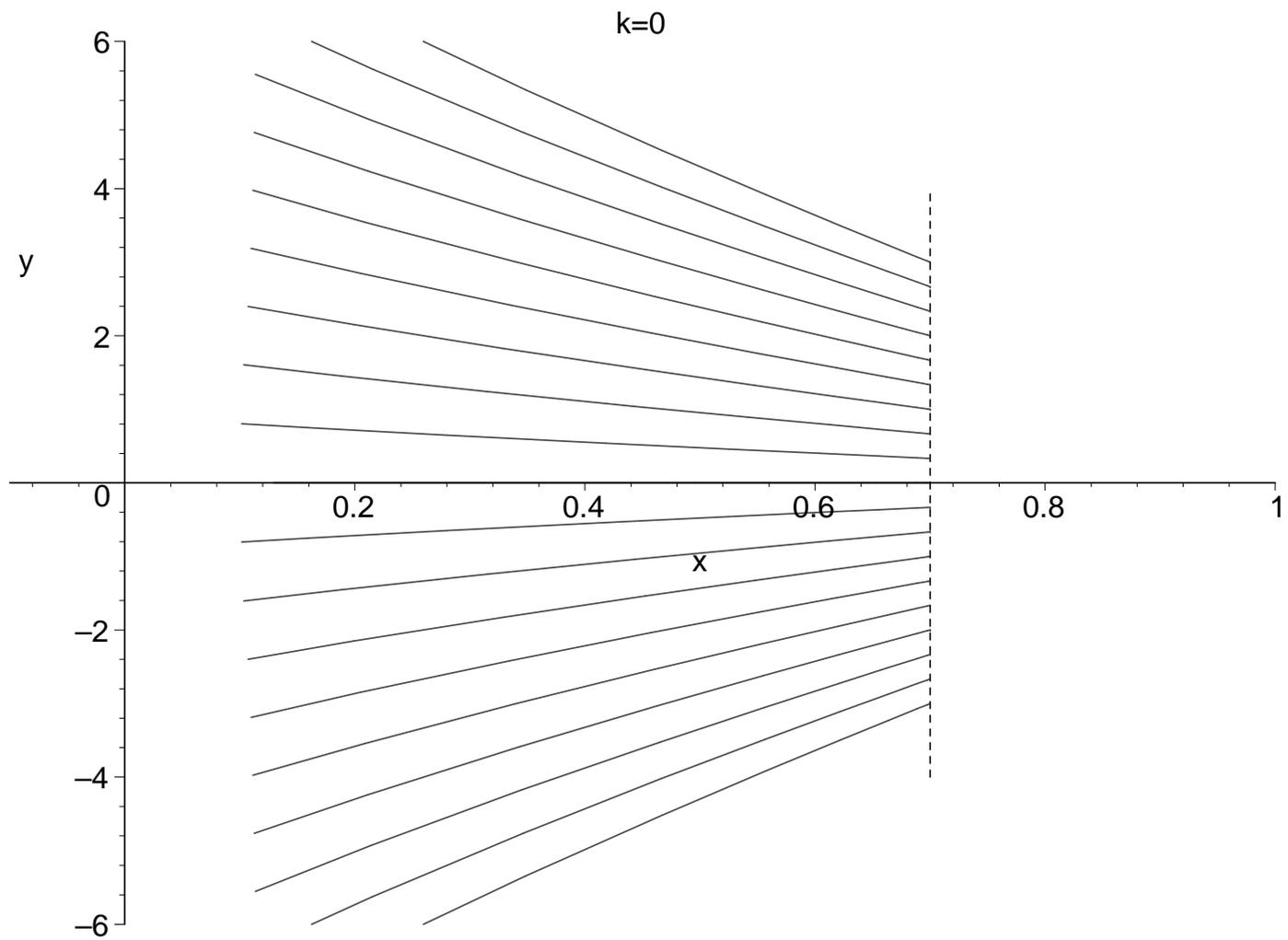
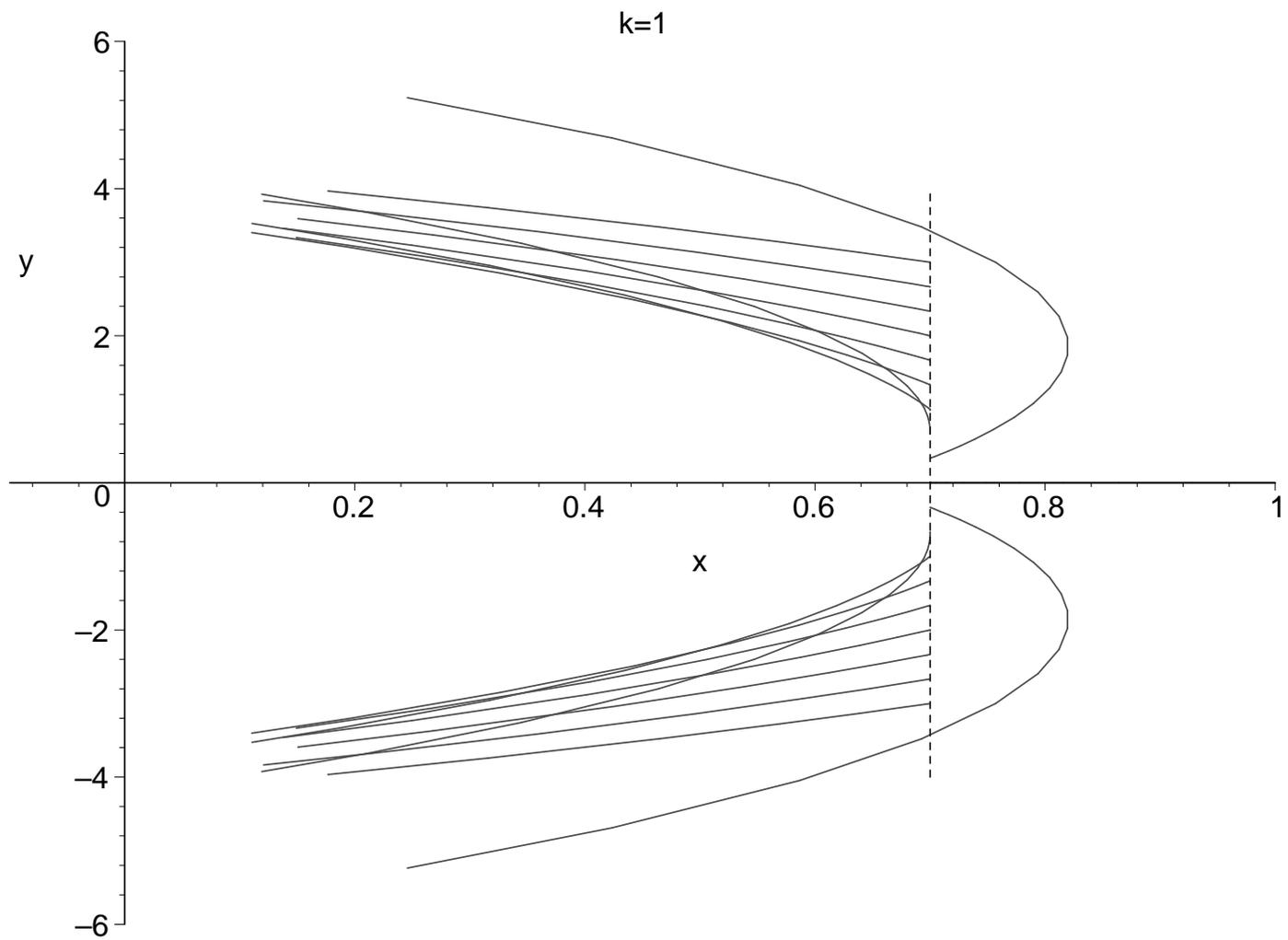
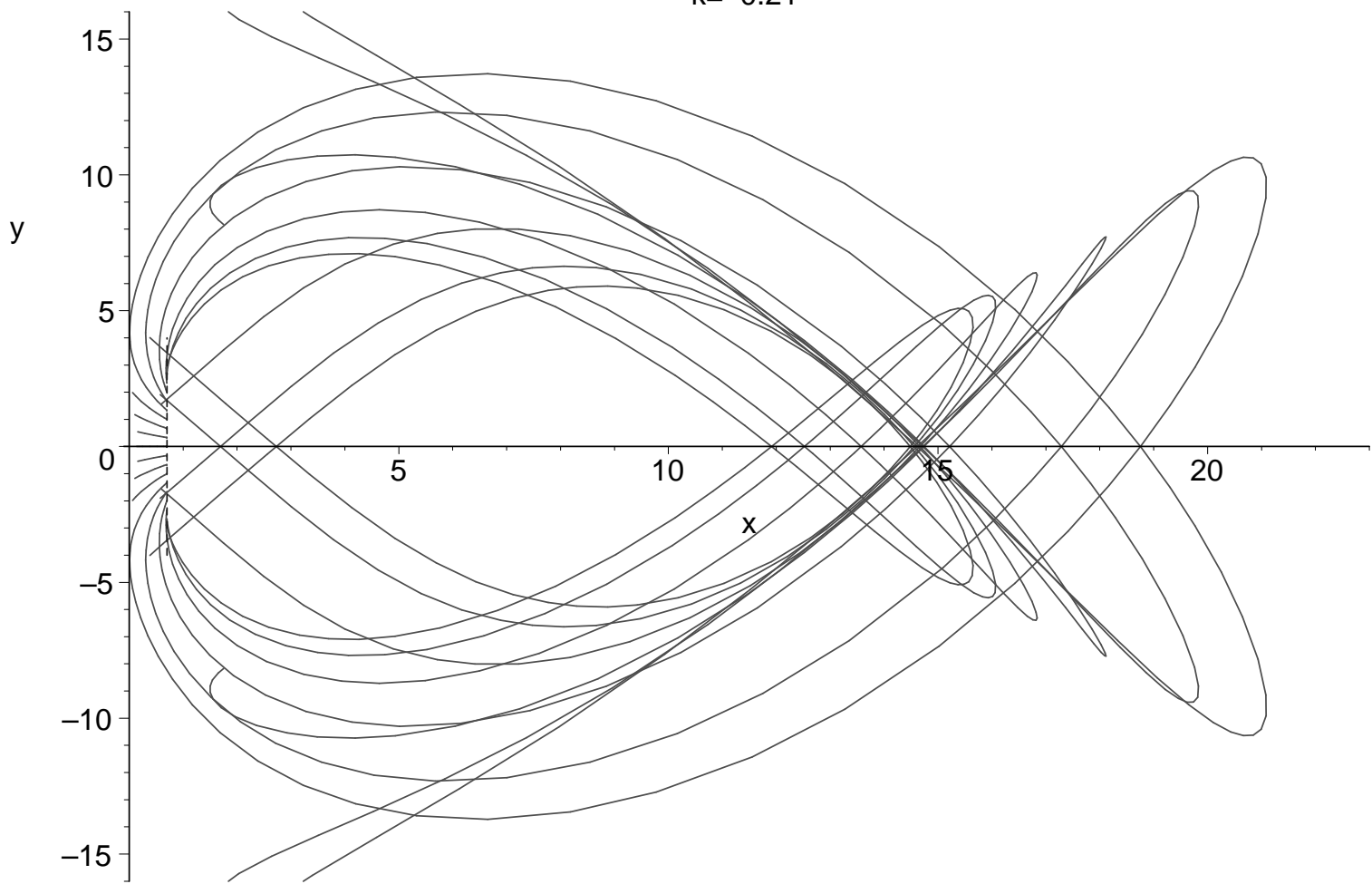
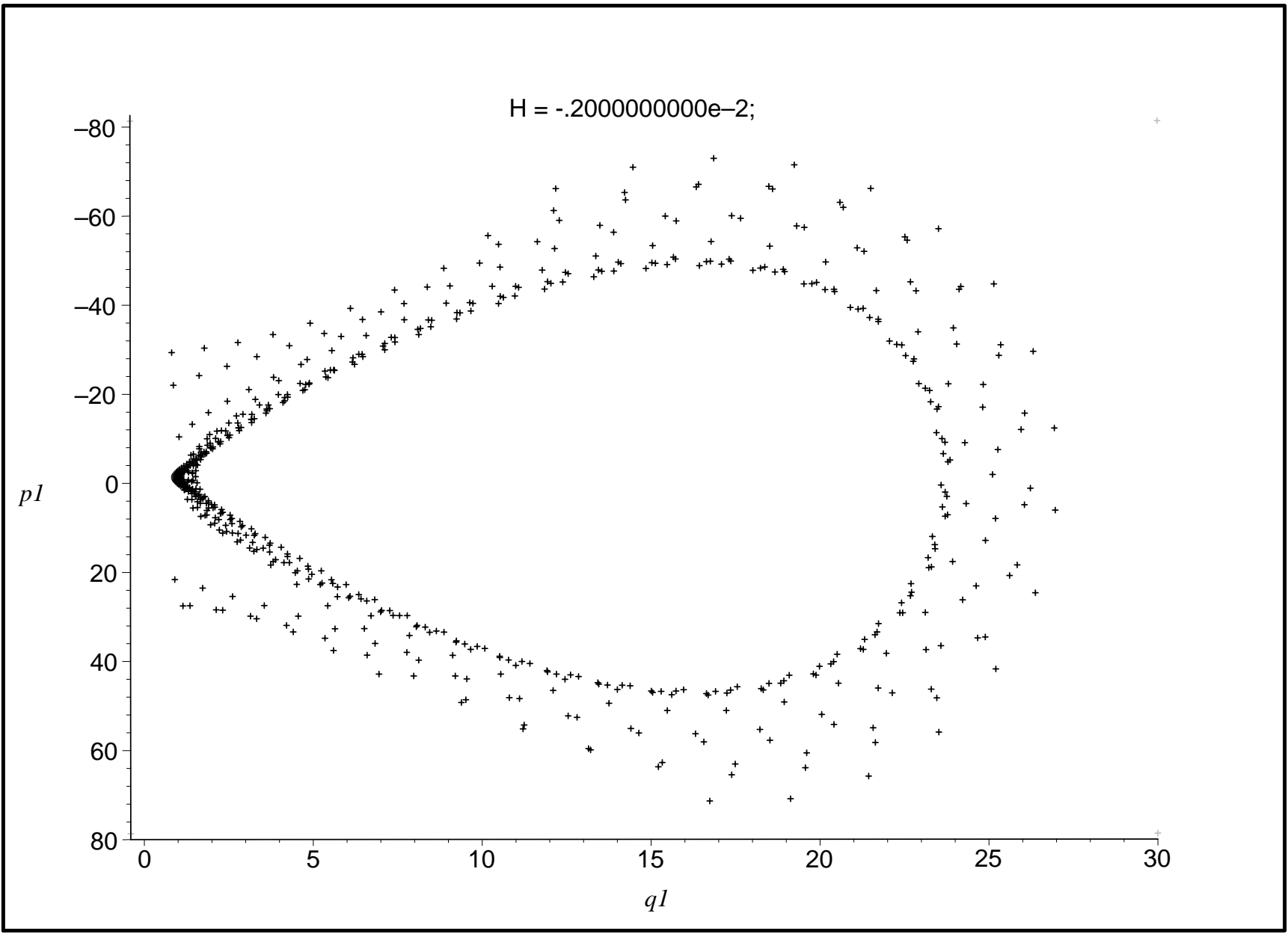
\end{document}